\documentclass[12pt]{article}
\newcommand{\be}{\begin{equation}}
\newcommand{\ee}{\end{equation}}
\newcommand{\ben}{\begin{equation*}}
\newcommand{\een}{\end{equation*}}
\newcommand{\ar}{\begin{array}}
\newcommand{\arn}{\end{array}}

\newcommand{\vk}{\vec{k}}

\newcommand{\q}{\vec{q}}
\newcommand{\qs}{\vec{q}^{\;2}}
\newcommand{\qp}{\vec{q}^{\;\prime}}

\newcommand{\x}{\vec{r}}
\newcommand{\xs}{\vec{r}^{\;2}}
\newcommand{\xp}{\vec{r}^{\;\prime}}

\def\pnot{\mbox{${\not{\hbox{\kern-3.0pt$p$}}}$}}
\def\qnot{\mbox{${\not{\hbox{\kern-2.0pt$q$}}}$}}
\def\enot{\mbox{${\not{\hbox{\kern-2.0pt$e$}}}$}}
\def\knot{\mbox{${\not{\hbox{\kern-2.0pt$k$}}}$}}
  \def\fun#1#2{\lower3.6pt\vbox{\baselineskip0pt\lineskip.9pt\ialign
{$\mathsurround=0pt#1\hfil##\hfil$\crcr#2\crcr\sim\crcr}}}

\begin{document}

\begin{titlepage}

\begin{center}
{\bf The dipole form of the quark part of the BFKL kernel $^{~\ast}$}
\end{center}

\vskip 0.5cm

\centerline{V.S.~Fadin$^{a\,\dag}$, R.~Fiore$^{b\,\ddag}$, A.~Papa$^{b\,\dag\dag}$}

\vskip .6cm

\centerline{\sl $^{a}$ Budker Institute of Nuclear Physics, 630090 Novosibirsk, Russia}
\centerline{\sl Novosibirsk State University, 630090 Novosibirsk, Russia}
\centerline{\sl $^{b}$ Dipartimento di Fisica, Universit\`a della Calabria,}
\centerline{\sl Istituto Nazionale di Fisica Nucleare, Gruppo collegato di Cosenza,}
\centerline{\sl Arcavacata di Rende, I-87036 Cosenza, Italy}

\vskip 2cm

\begin{abstract}
The dipole form of the ``Abelian'' part of the massless quark
contribution to the BFKL kernel is found in the coordinate
representation by direct transfer from the momentum representation
where the contribution was calculated before. It coincides with the
corresponding part of the quark contribution to the dipole kernel
calculated recently by Balitsky and is conformal invariant.
\end{abstract}


\vfill \hrule \vskip.3cm \noindent $^{\ast}${\it Work supported
in part by the Russian Fund of Basic Researches and in part by
Ministero Italiano dell'Istruzione, dell'Universit\`a e della
Ricerca.} \vfill $
\begin{array}{ll} ^{\dag}\mbox{{\it e-mail address:}} &
\mbox{FADIN@INP.NSK.SU}\\
^{\ddag}\mbox{{\it e-mail address:}} &
\mbox{FIORE@CS.INFN.IT}\\
^{\dag\dag}\mbox{{\it e-mail address:}} &
\mbox{PAPA@CS.INFN.IT}\\
\end{array}
$

\end{titlepage}

\vfill \eject

\section{Introduction}

The BFKL approach~\cite{BFKL} for the theoretical description of
high-energy  processes is well developed now in the
next-to-leading approximation (NLA). In particular, the kernel of
the BFKL equation is known in the next-to-leading order (NLO) not
only for the forward scattering~\cite{FL98}, i.e. for $t=0$ and
the color singlet in the $t$-channel, but also for arbitrary
fixed (not growing with energy) momentum transfer $t$ and any
possible color state in the $t$-channel~\cite{FFP99,FG00,FF05}.
The color singlet representation is certainly the most important
for phenomenological applications, since it is relevant for
physical processes, where the colliding particles are colorless,
although from the theoretical point of view the color octet
representation seems even more important because of the gluon
Reggeization. The Reggeized gluon occurs to be the primary Reggeon
in the high-energy QCD, which can be reformulated in terms of a
gauge-invariant effective field theory for the Reggeized gluon
interactions~\cite{Lipatov:1995pn}.

All the results mentioned above for the BFKL kernel have been
obtained so far in the momentum representation. However,
considering the singlet BFKL kernel in the coordinate
representation in the transverse space may have several
theoretical benefits.

Indeed, the famous property of conformal invariance of the BFKL
equation in the leading approximation~\cite{Lipatov:1985uk} is
connected with the coordinate representation. This property is
extremely important for finding solutions of the equation and it
is therefore very useful to investigate the conformal properties
of the BFKL kernel in the NLO. Here, an evident source of the
breaking of conformal invariance is the renormalization, but it is
important to know if this is the only source of violation. If so,
one can rely on the conformal invariance of the NLO BFKL in
supersymmetric extensions of QCD.

Moreover, it is in the coordinate representation that the color
dipole approach to high-energy scattering~\cite{dipole}, very
popular now, is formulated. Beside giving a clear physical picture
of the high-energy processes, the color dipole approach can be
naturally extended from the regime of low parton densities to the
saturation regime~\cite{GLR83}, where the evolution equations of
parton densities with energy become nonlinear. In general, there
is an infinite hierarchy of coupled equations~\cite{Balitsky,CGC}.
In the simplest case of a large nucleus as target, this set of
equations is reduced to the BK (Balitsky-Kovchegov)
equation~\cite{Balitsky}.

The theoretical description of small-$x$ processes could take
advantage from a clear understanding of the relation between the
BFKL and the color dipole approaches. It is
affirmed~\cite{dipole,Balitsky}, that in the linear regime the
color dipole gives the same results as the BFKL approach for the
color singlet channel. However, a full insight into this
relation and its extension to the NLO requires to bridge the
gap between the different formulations adopted by the two
approaches: the color dipole approach uses the coordinate
representation in the transverse space, whereas the BFKL approach
was originally formulated in the momentum one.

Before the advent of the dipole approach, the leading order color
singlet BFKL kernel has been investigated in the coordinate
representation in detail in Ref.~\cite{Lipatov:1985uk}. More
recently, the relation between BFKL and color dipole was analyzed
in the leading order in Ref.~\cite{Bartels:2004ef}. The extension
of the analysis to the NLO has started in the last few months. In
Ref.~\cite{Kovchegov:2006wf} the quark contribution in the color
dipole approach at large number of colors $N_c$ has been
transferred from the coordinate to the momentum representation and
it has been verified that the resulting contribution to the NLO
Pomeron intercept agrees with the well-known result of
Ref.~\cite{FL98}. In our previous paper~\cite{Fadin:2006ha} we
have  transformed  the ``non-Abelian'' (leading in $N_c$) part of
the quark contribution to the non-forward BFKL kernel from the
momentum representation where it was calculated
before~\cite{FFP99} to the coordinate one and have found that its
dipole form is in accord with the result obtained recently in
Ref.~\cite{Balitsky:2006wa} by direct calculation of the quark
contribution to the dipole kernel in the coordinate
representation.  In this paper we consider the ``Abelian''
(suppressed by $1/N_c^2$) part of the quark contribution to the
non-forward  BFKL kernel~\cite{FFP99}.

The paper is organized as follows: in Section~\ref{sec:notation}
we give the basic definitions, fix our notations and recall the
main results of our previous paper~\cite{Fadin:2006ha}; in
Section~\ref{sec:abel} we describe the procedure to transfer the
``Abelian'' part of the NLO quark contribution from the momentum
to the coordinate representation and present our result; in
Section~\ref{sec:conclusion} we draw our conclusions.

\section{Basic definitions and notation}
\label{sec:notation}

We use the same notation as in Ref.~\cite{Fadin:2006ha}: $\qp_i$
and $\q_i$, $i=1,2$, represent the transverse momenta of Reggeons
in initial and final $t$-channel states, while $\xp_i$ and $\x_i$
are the corresponding conjugate coordinates. The state
normalization is
\be \label{normalization}\langle
\q|\qp\rangle=\delta(\q-\qp)\;, \;\;\;\;\; \langle
\x|\xp\rangle=\delta(\x-\xp)\;,
\ee
so that
\be
\langle\x|\q\rangle=\frac{e^{i\q\;\x}}{(2\pi)^{1+\epsilon}}\;,
\ee
where $\epsilon=(D-4)/2$; $D-2$ is the dimension of the transverse
space and is taken different from $2$ for the regularization of
divergences. Note that in our papers previous to Ref.~\cite{Fadin:2006ha}
we denoted the initial (final) momenta as $\q_1$ and $-\qp_1$ ($\q_2$ and
$-\qp_2$) and used the normalization
$\langle\q|\qp\rangle=\qs\delta(\q-\qp)$. We will use also the
notation $\q=\q_1+\q_2,\;\;\qp=\qp_1+\qp_2;
\;\;\vk=\q_1-\qp_1=\qp_2-\q_2$. The BFKL kernel in the operator
form is written as
\begin{equation}\label{operator of the BFKL kernel}
\hat{\cal K}=\hat{\omega}_1+\hat{\omega}_2+ \hat{\cal K}_r\;,
\end{equation}
where \be\label{trajectory ff} \langle \q_{i}|\hat{\omega}_i|
\qp_{i}\rangle=\delta(\q_i-\qp_i)\omega(-\qs_i )\;, \ee with
$\omega(t)$ the gluon Regge trajectory, and $\hat{\cal K}_r$
represents real particle production in Reggeon collisions. The
$s$-channel discontinuities of scattering amplitudes for the
processes $A+B\rightarrow A^\prime +B^\prime$ have the form
\be\label{discontinuity representation}
-4i(2\pi)^{D-2}\delta(\q_A-\q_B)\mbox{disc}_s{\cal
A}_{AB}^{A'B'}=\langle A^\prime \bar A|e^{Y\hat{\cal
K}}\frac{1}{\hat{\q}^{\;2}_1\hat{\q}^{\;2}_2}|\bar B^\prime
B\rangle\;. \ee In this equation $Y=\ln(s/s_0)$, $s_0$ is an
appropriate energy scale, $\;\;q_A=p_{A'}-p_A,\;\;q_B=p_B-p_{B'}$,
and \be\label{kernel ff} \langle\q_{1},\q_{2}|\hat{\cal
K}|\qp_{1},\qp_{2}\rangle =\delta(\q-\qp)\frac{1}{\qs_1\qs_2}{\cal
K}(\q_1,\qp_1;\q)\;, \ee \be\label{impact BB} \langle
\q_{1},\q_{2}|\bar B^\prime B\rangle=4p_B^-\delta(\q_{B}-\q_{1}
-\q_{2}){\Phi}_{B'B}(\q_{1},\q_{2})\;, \ee \be\label{impact AA}
\langle A^\prime \bar A|\q_{1},\q_{2}\rangle=
4p_A^+\delta(\q_{A}-\q_{1}-\q_{2}){\Phi}_{A'A}(\q_{1},\q_{2})\;,
\ee where $p^{\pm}=(p_0\pm p_z)/\sqrt 2$; the kernel ${\cal
K}(\q_1,\qp_1;\q)$ and the impact factors $\Phi$ are expressed
through the Reggeon vertices according to Ref.~\cite{FF98}. Note
that the appearance of the factors
$(\hat{\q}^{\;2}_1\hat{\q}^{\;2}_2)^{-1}$ in (\ref{discontinuity
representation}) and  $(\qs_1\qs_2)^{-1}$ in (\ref{kernel ff})
cannot be explained by a change of the
normalization~(\ref{normalization}). We have used a freedom in
the definition of the kernel. In fact, one can change the form of the
kernel performing the transformation
\begin{equation}\label{kernel transformation}
\hat{\cal K}\rightarrow \hat{\cal O}^{-1}\hat{\cal K}\hat{\cal
O}~,\;\; \langle A^\prime \bar A|\rightarrow \langle A^\prime \bar
A|\hat{\cal O}~,
\;\;\frac{1}{\hat{\q}^{\;2}_1\hat{\q}^{\;2}_2}|\bar B^\prime
B\rangle \rightarrow {\hat{\cal
O}^{-1}}\frac{1}{\hat{\q}^{\;2}_1\hat{\q}^{\;2}_2}|\bar B^\prime
B\rangle\;,
\end{equation}
which does not change the discontinuity (\ref{discontinuity
representation}). In (\ref{kernel transformation}) $\hat{\cal O}$
is an arbitrary nonsingular operator. The kernel $\hat{\cal K}$ in
(\ref{kernel ff}) is related with the one defined in
Ref.~\cite{FF98} by such transformation with $\hat{\cal
O}=(\hat{\q}^{\;2}_1\hat{\q}^{\;2}_2)^{1/2}$. The reason for this
choice is that in the leading order the kernel which is conformal
invariant and is simply related to the dipole kernel is not the
kernel defined in Ref.~\cite{FF98}, but just the kernel
 $\hat{\cal K}$ in (\ref{kernel ff})~\cite{Lipatov:1985uk,Bartels:2004ef}.
Note that after the choice of  $\hat{\cal O}$ in the leading order,
transformations with $\hat{\cal O}=1-\hat O$, where $\hat O\sim
g^2$,  are still possible. At the NLO after such transformation we
get
\begin{equation}\label{transformation at NLO}
\hat{{\cal K}} \rightarrow\hat{{\cal K}}-[\hat{{\cal
K}}^{(B)},\hat O]~,
\end{equation}
where $\hat{{\cal K}}^{(B)}$ is the leading order kernel.

In Ref.~\cite{Fadin:2006ha} we performed the transformation to the
coordinate representation of the ``non-Abelian'' part of the quark
contribution to the kernel $\langle\q_{1},\q_{2}|\hat{\cal
K}|\qp_{1},\qp_{2}\rangle$ (\ref{kernel ff}) which was found
in~\cite{FFP99}.  The transformation was performed in the most
general way: at arbitrary $D$ and for the case of arbitrary impact
factors. In the case of scattering of colorless objects besides
the freedom of definition of the kernel discussed above there is
an additional freedom related  with the ``gauge invariance"
(vanishing at zero Reggeized gluon momenta) of the impact factors
\cite{Lipatov:1985uk,Bartels:2004ef}. In this case  the kernel
$\langle \x_1\x_2|\hat{{\cal K}}|\xp_1\xp_2\rangle$ can be written
in the dipole form (see below). Because of the possibility of
transformations (\ref{transformation at NLO}) the dipole form is
not unique. We  transferred to the  dipole form the
``non-Abelian'' part  of the kernel~\cite{FFP99} and found that,
after the transformation (\ref{transformation at NLO}) with a suitable
operator $\hat O$, it  agrees
with the result obtained recently in Ref.~\cite{Balitsky:2006wa}
by direct calculation of the quark contribution to the dipole
kernel in the coordinate representation. Here we consider also the
``Abelian'' part of the kernel and find its dipole form.

\section{The ``Abelian'' part of the quark contribution}
\label{sec:abel}

The ``Abelian'' part of the quark contribution to the NLO kernel
is suppressed by the factor $1/N_c^2$ in comparison with the
``non-Abelian'' one~\cite{FFP99}. The gluon trajectory has only
leading $N_c$ contribution. Therefore, the ``Abelian''
contribution comes only from the real part of the kernel. This
part contains neither ultraviolet nor infrared singularities and
therefore it does not require regularization nor renormalization.
Therefore we can use from the beginning physical space-time
dimension $D=4$ and the renormalized coupling constant
$\alpha_s(\mu)$. According to  Eqs.~(37), (38), (45) and (49) of
Ref.~\cite{FFP99} the ``Abelian'' contribution in the momentum
representation can be written as
\begin{equation}
\langle \q_1,\q_2|\hat{{\cal K}}^a|\qp_1,\qp_2\rangle =
\delta(\q-\qp)\frac{\alpha_s^2(\mu)n_f}{(2\pi
)^{2}N_c}\frac{-2}{\qs_1\qs_2}\int_0^1 dx\int
\frac{d^2k_1}{(2\pi)^2}F(\q_1,\q_2; \vk_1,\vk_2)\;,  \label{kernel
through F}
\end{equation}
with
\[
F(\q_1,\q_2; \vk_1,\vk_2)= x(1-x)\left( \frac{2(
\vec{q}_{1}\vec{k}_{1})-\vec{q}_{1}^{\:2}}{\sigma_{11}}+\frac{2(\vec{q}_{1
} \vec{k}_{2})-\vec{q}_{1}^{\:2}}{\sigma_{21}}\right) \] \[ \times
\left( \frac{2(\vec{q}_{2}\vec{k}_{1})+\vec{q}
_{2}^{\:2}}{\sigma_{12}}+\frac{2(\vec{q}_{2}
\vec{k}_{2})+\vec{q}_{2}^{\:2}}{\sigma_{22}}\right) + \frac{x
\vec{q}^{\:2}(2(\vec{q}_{1}\vec{k}_{1})-\vec{q}_{1}^{\:2})}{2\sigma
_{11}}\left( \frac{1}{\sigma_{22}}-\frac{1}{\sigma_{12}}\right) \]
\[ +\frac{x\vec{q}^{\:2}(2(\vec{q}_{2}\vec{k}_{1})+\vec{q} _{2}^{
\:2})}{2\sigma_{12}}\left(
\frac{1}{\sigma_{11}}-\frac{1}{\sigma_{21}}\right)
+\frac{1}{\sigma_{11}\sigma_{22}}\biggl(
-2(\vec{q}_{1}\vec{k}_{1})(\vec{q} _{2}\vec{q}_{2}^{\:\prime
~})\biggr. \] \begin{equation} \biggl.
-2(\vec{q}_{2}\vec{k}_{1})(\vec{q}_{1}\vec{q}
_{1}^{\:\prime})+(\vec{q}_{2}^{
\:2}-\vec{q}_{1}^{\:2})(\vec{k}_{1}\vec{k})+\vec{q}
_{1}^{\:2}\vec{q}_{2}^{\:\prime
\:2}-\frac{{\vec{k}}_{{}}^{2}\vec{q}_{{}}^{\:2}}{ 2}\biggr) \;,
\label{F in p space} \end{equation}
where $\vk_1+\vk_2=\vk=\q_1-\qp_1=\qp_2-\q_2$,
\[
\sigma_{11}=(\vk_1-x\q_1)^2+x(1-x)\qs_1,
\;\;\sigma_{21}=(\vk_2-(1-x)\q_1)^2+x(1-x)\qs_1\;,
\]
\begin{equation} \sigma_{12}=(\vk_1+x\q_2)^2+x(1-x)\qs_2,
\;\;\sigma_{22}=(\vk_2+(1-x)\q_2)^2+x(1-x)\qs_2\;. \label{sigma}
\end{equation}
It is easy to see that $F(\q_1,\q_2; \vk_1,\vk_2)$ vanishes
when any of the $\vec{q}_{i}$'s or $\vec{q}_{i}^{\:\prime}$'s
tends to zero. The vanishing of $\langle \q_1,\q_2|\hat{{\cal
K}}^a|\qp_1,\qp_2\rangle$ at $\vec{q}_{i}^{\:\prime }=0$ together
with the ``gauge invariance'' property of the impact factors of
colorless ``projectiles" ${\Phi}_{A'A}(0, \q)={\Phi}_{A'A}(\q,0)$
permits to omit in $\langle \x_1\x_2|\hat{{\cal
K}}^{a}|\xp_1\xp_2\rangle$  terms which do not depend either on
$\x_1$ or on  $\x_2$. Moreover, it permits to change in~(\ref{discontinuity representation})
the ``target" impact factors so that they acquire the ``dipole"  property
\begin{equation}\label{input}
\langle \x, \x|({\hat{\q}^{\;2}_1\hat{\q}^{\;2}_2})^{-1}|\bar
B^\prime B\rangle_d=0\;,
\end{equation}
(see Ref.~\cite{Fadin:2006ha} for details). Thereafter, terms
proportional to $\delta(\xp_1-\xp_2)$ in $\langle
\x_1\x_2|\hat{{\cal K}}^{a}|\xp_1\xp_2\rangle$ can also be
omitted, assuming that the remaining part $\hat{{\cal K}}^{a}_d$
conserves the ``dipole"  property. We call this part the dipole
form of the BFKL kernel.

With our normalizations, $\hat{\cal K}^{a}$ in the coordinate
representation is given by
\begin{equation}\label{real kernel through f}
\langle \x_1\x_2|\hat{{\cal K}}^{a}|\xp_1\xp_2\rangle =-2
\frac{\alpha_s^2(\mu)n_f}{(2\pi )^{4}N_c}\int_0^1 dx\int
\frac{d^2q_1}{2\pi}\frac{d^2q_2}{2\pi}\frac{d^2k_1}{2\pi}\frac{d^2k_2}{2\pi}
\frac{F(\q_1,\q_2;\vk_1,\vk_2)}{\qs_1\qs_2}
\end{equation}
\[
\times
e^{i[\q_1(\x_1-\xp_1)+\q_2(\x_2-\xp_2)+\vk(\xp_1-\xp_2)]}\;.
\]
If we restrict ourselves to the dipole form, then, omitting the
terms with $\delta(\xp_1-\xp_2)$, we can change the integrand as follows:
\[
F(\q_1,\q_2; \vk_1,\vk_2)\rightarrow
\frac{1}{\sigma_{11}\sigma_{22}}\Biggl[ 2x(1-x)(2
\vec{q}_{1}\vec{k}_{1}-\vec{q}_{1}^{\:2})(2\vec{q}_{2}\vec{k}_{2}+\vec{q}
_{2}^{\:2})+\frac{\vec{q}^{\:2}}{2}\left(x
(2\vec{q}_{1}\vec{k}_{1}-\vec{q}_{1}^{\:2})\right.
\]
\begin{equation}
\left.-(1-x)(2\vec{q}_{2}\vec{k}_{2}+\vec{q}
_{2}^{\:2})\right) -2(\vec{q}_{1}\vec{k}_{1})(\vec{q}
_{2}\vec{q}_{2}^{\:\prime
~})-2(\vec{q}_{2}\vec{k}_{1})(\vec{q}_{1}\vec{q}
_{1}^{\:\prime})+(\vec{q}_{2}^{
\:2}-\vec{q}_{1}^{\:2})(\vec{k}_{1}\vec{k})+\vec{q}
_{1}^{\:2}\vec{q}_{2}^{\:\prime
\:2}-\frac{{\vec{k}}_{{}}^{2}\vec{q}_{{}}^{\:2}}{ 2}\Biggr] ~.
\label{simplified F}
\end{equation}
Note that the ``Abelian'' part of the quark contribution is given by
the ``box'' and ``cross-box'' diagrams; the ``box''  diagrams give only
terms proportional to $\delta(\xp_1-\xp_2)$, whereas all contribution
of ``cross-box'' diagrams is retained in~(\ref{simplified F}).
It is well known that in the momentum representation this contribution is
the most complicated one. On the contrary, in the coordinate representation
it is very simple.
The integrals can be calculated quite easily. The integrations
over $\vk_1, \q_1$ and $\vk_2, \q_2$ can be done independently. It
is convenient to introduce the new variables $\vec l_1$ and $\vec
l_2$,
\begin{equation}\label{l i}
\vk_1= \vec l_1+x\q_1,\;\;\vk_2= \vec
l_2 -(1-x)\q_2,\;\;\sigma_{11}=\vec l_1^{\: 2}+x(1-x)\qs_1,
\;\;\sigma_{22}=\vec l_2^{\: 2}+x(1-x)\qs_2\;.
\end{equation}
Omitting the terms with $\delta(\xp_1-\xp_2)$, we obtain in
terms of these variables
\[
 F(\q_1,\q_2;
\vk_1,\vk_2)\rightarrow \frac{1}{\sigma_{11}\sigma_{22}}\Biggl[
-8x^2(1-x)^2\qs_1\qs_2-4x(1-x)(1-2x)\left(
\qs_1(\vec{q}_{2}\vec{l}_{2})+\qs_2(\vec{q}_{1}\vec{l}_{1})\right)
\]
\begin{equation}\label{simplified integrand}
+2\left(4x(1-x)-1\right)(\vec{q}_{1}\vec{l}_{1})(\vec{q}_{2}\vec{l}_{2})
+2(\vec{q}_{2}\vec{l}_{1})(\vec{q}_{1}\vec{l}_{2})
-2(\vec{q}_{1}\vec{q}_{2})(\vec{l}_{1}\vec{l}_{2})\Biggr]\;.
\end{equation}
The only integrals we need are
\[
\int
\frac{d^2q}{2\pi}\frac{d^2l}{2\pi}\frac{e^{i\q\x+i\vec{l}\vec\rho}}
{\vec{l}^{\;2}+x(1-x)\qs}=\frac{1}{\xs+x(1-x)\vec\rho^{\;2}}\;,
\]
\begin{equation}\label{integrals}
\int\frac{d^2q}{2\pi}\frac{d^2l}{2\pi}\frac{q_il_j}{\qs}
\frac{e^{i\q\x+i\vec{l}\vec\rho}}{\vec{l}^{\;2}+x(1-x)\qs}
=\frac{-r_i\rho_j}{\vec\rho^{\;2}(\xs+x(1-x)\vec\rho^{\;2})}\;.
\end{equation}
We obtain
\[
\langle \x_1\x_2|\hat{{\cal K}}^{a}|\xp_1\xp_2\rangle =
\frac{\alpha_s^2(\mu)n_f}{4\pi^{4}N_c}\int_0^1
\frac{dx}{d_1d_2(\xp_1-\xp_2)^4}\Biggl[\left((\x_1-\xp_2)^2
-(\x_1-\xp_1)^2\right)\left((\x_2-\xp_1)^2-(\x_2-\xp_2)^2\right)
\]
\[
\times x(1-x) +
(\xp_1-\xp_2)^2\Biggl((\xp_1-\xp_2)\Bigl(x(\x_2-\xp_2)-(1-x)(\x_1-\xp_1)
\Bigr) +(\x_1-\xp_1)(\x_2-\xp_2)\Biggr)\Biggr]\; \]
\begin{equation}\label{kernel before integration over x} + \mbox{ terms
with } \delta(\xp_1-\xp_2)\;. \end{equation} Here
$d_1=x(\x_1-\xp_2)^2+(1-x)(\x_1-\xp_1)^2$ and
$d_2=x(\x_2-\xp_2)^2+(1-x)(\x_2-\xp_1)^2$. The subsequent
integration over $x$ is elementary. We need only the integrals
\[
\int_0^1 \frac{dx}{d_1d_2}=\frac{L_+}{d}\;, \] \[ \int_0^1
\frac{xdx}{d_1d_2}=\frac{1}{d}\left(\frac{c_2}{b_2}l_2
-\frac{a_1}{b_1}l_1\right)\;, \] \[ \int_0^1
\frac{(1-x)dx}{d_1d_2}=\frac{1}{d}\left(\frac{c_1}{b_1}l_1
-\frac{a_2}{b_2}l_2\right)\;,
\]
\begin{equation}\label{integration over x} \int_0^1
\frac{x(1-x)dx}{d_1d_2}=\frac{1}{b_1b_2}
\Biggr[1-\frac{1}{d}\left(\frac{a_1c_1b_2}{b_1}l_1+
\frac{a_2c_2b_1}{b_2}l_2\right)\Biggr]\;,
\end{equation}
where
\[
a_1=(\x_1-\xp_1)^2\;, \;\;\; a_2=(\x_2-\xp_2)^2\;, \;\;\;
c_1=(\x_1-\xp_2)^2\;, \;\;\; c_2=(\x_2-\xp_1)^2\;,
\;\;\;b_i=c_i-a_i\;,
\]
\begin{equation}
d=c_1c_2-a_1a_2\;, \;\;\;
l_i=\ln\left(\frac{c_i}{a_i}\right)\;, \;\;\;L_+=l_1+l_2\;.
\end{equation}
Calculating the integrals with the help of these
formulas, we obtain, up to terms with $\delta(\xp_1-\xp_2)$,
\[
\langle \x_1\x_2|\hat{{\cal K}}^{a}|\xp_1\xp_2\rangle =
\frac{\alpha_s^2(\mu)n_f}{4\pi^{4}N_c}\frac{1}{(\xp_1-\xp_2)^4}
\Biggl[\left(\frac{c_1c_2+a_1a_2-(\x_1-\x_2)^2(\xp_1-\xp_2)^2}{2d}L_+
-1\right)
\]
\begin{equation}\label{kernel after integration over x}
-\left(\frac{c_1+a_1-(\xp_1-\xp_2)^2}{2b_1}l_1-1\right)
-\left(\frac{c_2+a_2-(\xp_1-\xp_2)^2}{2b_2}l_2-1\right)\Biggr]\;.
\end{equation}
The terms in the last line do not depend either on $\x_1$ or
$\x_2$ and can therefore be omitted due to the gauge invariance of
impact factors. The remaining part turns into zero at $\x_1=\x_2$
(conserves the ``dipole" property),  so that it represents the
dipole form of the ``Abelian" part of the quark contribution to
the BFKL kernel. It coincides with the corresponding part of the
quark contribution to the dipole kernel calculated recently in
Ref.~\cite{Balitsky:2006wa} and is evidently conformal invariant.

\section{Conclusion}
\label{sec:conclusion}

The coordinate representation of the BFKL kernel is extremely
interesting, because it gives the possibility to understand its
conformal properties and the relation between the BFKL and the
color dipole approaches. Joining the results of
Ref.~\cite{Fadin:2006ha} and of the present paper, we have
obtained the quark contribution to the BFKL kernel in the
next-to-leading order in the coordinate representation in the
transverse space, by transformation from the momentum
representation, in which the BFKL kernel was known before. For
scattering of colorless objects, due to the ``gauge invariance" of
the impact factors,  the kernel in the coordinate representation
can be written in the dipole form. We have found that  the dipole
form of the quark contribution  agrees with the result obtained
recently in Ref.~\cite{Balitsky:2006wa} by direct calculation of
the quark contribution to the dipole kernel in the coordinate
representation. The agreement is reached after some
transformations of the original BFKL kernel  which do not change
the scattering amplitudes, being supplemented by the corresponding
transformation of the impact factors of colliding particles.

As for the ``Abelian'' part of the kernel, which is the subject of
the present paper, it turns out that its dipole form is quite
simple as compared with the very complicated form in the momentum
representation. Actually, up to a coefficient, this part
coincides~\cite{{FFP99}} with the kernel for the QED
Pomeron~\cite{Non-forward Pomeron in QED}, so that its complexity
in the momentum representation has been known for a long time.
Moreover, the dipole form of the ``Abelian'' part of the kernel is
conformal invariant. It could be especially interesting for the
QED Pomeron. However, one has to remember that in QED the use of
the dipole form is limited to scattering of neutral objects, as
well as that the conformal invariance is broken by masses.

\vspace{1.0cm} \noindent {\Large \bf Acknowledgment}
\vspace{0.5cm}

A.P. thanks A. Sabio Vera for several useful discussions.

\end{document}